\documentclass[twocolumn]{revtex4}
\usepackage{hyperref}
\usepackage{amsmath,amssymb}
\usepackage{epsfig}
\usepackage{graphicx}
\usepackage{dcolumn}

\DeclareMathOperator{\arccot}{arccot}
\begin{document}

\title{Guiding of vortices and ratchet effect \\ in superconducting films
with asymmetric pinning potential}

\author{Valerij A. Shklovskij$^{1,2}$ and Vladimir V.
Sosedkin$^{2}$}

\address
{$^1$ Institute of Theoretical Physics, National Science
Center-Kharkov Institute of Physics and Technology, 61108,
Kharkov, Ukraine \\
$^2$Kharkov National University, Physical Department, 61077,
Kharkov, Ukraine}

\begin{abstract}
Two-dimensional vortex dynamics in a ratchet washboard planar
pinning potential (PPP) in the presence of thermal fluctuations is
considered on the basis of a
 Fokker-Planck equation. Explicit expressions for two new nonlinear anisotropic
 voltages (longitudinal and transverse with respect to the current direction) are
 derived and analyzed. The physical origin of these odd (with respect to
 magnetic field or transport current direction reversal) voltages is caused by the
  interplay between the even effect of vortex guiding and the ratchet asymmetry.
  Both new voltages are going to zero in the linear regimes of the vortex
  motion ( i.e. in the thermally activated flux flow (TAFF) and ohmic flux flow (FF)
  regimes) and have a bump-like current or temperature dependence in the
  vicinity of the highly nonlinear resistive transition from the TAFF to the FF.
\end{abstract}
\pacs{74.60.Ge, 74.72.Bk, 72.15.Gd, 74.25.Fy, 74.72.Bk, 75.15.Gd,}
\maketitle

\section{INTRODUCTION}
Now vortex ratchets, which exploit asymmetric vortex dynamics have
been attracting considerable attention
\cite{Reimann}-\cite{short}. The common feature of superconducting
ratchet systems is their rectifying property: the application of
the alternating current to a superconductor patterned with a
periodic asymmetric pinning potential can produce vortex motion
whose direction is determined only by the asymmetry of the
pattern.

Although considerable theoretical work exists \cite{Reimann}, only
few experiments have been realized. Recently a vortex lattice
ratchet effect has been investigated in Nb films sputtered on
arrays of nanometric Ni triangles, which produce the periodic
asymmetric pinning potential \cite{Villegas}. Similar effects were
also discussed for YBCO films with antidots \cite{Wordenweber}.
Earlier it has been proposed in \cite{Lee} how the ratchet effect
can be used to remove vortices from low-temperature
superconductors.

Unfortunately, a full temperature-dependent theoretical
description of the superconducting devices proposed in
\cite{Reimann}, is not available due to the complexity of the
two-dimensional periodic pinning potential in \cite{Reimann} used.
In particular, a theoretical explanation of the experimentally
available study of the vortex flow along the vortex channeling
directions in above-mentioned structures is a difficult problem.
Due to this reason we propose below to study experimental ratchet
properties of superconductors on the basis of a more simple
ratchet device for which exist a full theoretical description (at
least, in the single-vortex approximation) of its two-dimensional
vortex dynamics within the framework of a Fokker-Planck approach.

It is noticeable, that such a device has already been exploited
many years ago by Morrison and Rose in their experiments on
controlled asymmetric (as now we say ``ratchet``) surface pinning
in the superconducting-alloy films \cite{Morrison}. Recent
progress in fabrication of submicrometric structures with a
periodic ratchet modulation of their thickness by methods of
electron-beam lithography \cite{Martin} or molecular-beam epitaxy
on facetted substrates \cite{Huth} allows to prepare Nb films with
a similar well-controlled asymmetric washboard pinning structure.
Note also, that the main feature of similar structures is the
existence of well-defined \emph{guiding }of vortices along the
channels of the washboard pinning potential at relatively low
temperatures.

One of the first experimental observations of guided vortex motion
in the flux-flow regime was made by Niessen and Weijsenfeld still
in 1969 ~\cite{NW}. They studied guided vortex motion in the
cold-rolled sheets of a Nb-Ta alloy by measuring transverse
voltages of the pattern for different magnetic fields \emph{H},
transport current densities \emph{J}, temperatures \emph{T}, and
different angles $\alpha$ between the rolling and current
direction. The (\emph{H,J,T},$\alpha$)-dependences of the
cotangent of the angle $\beta$ between the average vortex velocity
$\langle \textbf{v}\rangle$ and the \textbf{j} direction were
presented. For the discussion, a simple theoretical model was
suggested, based on the assumption that vortex pinning and guiding
can be described in terms of an isotropic pinning force plus a
pinning force with a fixed direction which was perpendicular to
the rolling direction. The experimentally observed dependence of
the transverse and longitudinal voltages on the magnetic field
\emph{in the flux flow regime} as a function of the angle $\alpha$
was in agreement with this model. However, the dynamics of the
vortex that is moving transverse to the pinning channels has
substantively nonlinear behavior and cannot be entirely explained
within the flux-flow approach.

The \emph{nonlinear guiding} problem was exactly solved at first
only for washboard PPP within the framework of the two-dimensional
stochastic model of anisotropic pinning which takes into account
the vortex and the Hall viscosity coefficients and based on the
Fokker-Planck equation with a concrete form of the symmetric
pinning potential \cite{mawatari,ShSo}.

Rather simple formulas were derived in ~\cite{ShSo} for the
experimentally observable \emph{nonlinear} even(+) and odd(-)
(with respect to the magnetic field reversal) longitudinal and
transverse magnetoresistivities
$\rho_{\|,\perp}^\pm(j,t,\alpha,\epsilon)$ as functions of the
dimensionless transport current density $j,$ dimensionless
temperature $t,$ and relative volume fraction $0<\epsilon<1$
occupied by the parallel twin planes directed at an angle $\alpha$
with respect to the current direction. The
$\rho_{\|,\perp}^\pm$-formulas were presented in \cite{ShSo} as
linear combinations of the even and odd parts of the function
$\nu(j,t,\alpha,\epsilon)$ which can be considered as the
probability of overcoming the potential barrier of the pinning
channel; this made it possible to give a simple physical treatment
of the nonlinear regimes of vortex motion.

In addition to the appearance a of well known a relatively large
even transverse $\rho_\perp^+$ resistivity [9], generated by the
guiding of vortices along the channels of the washboard PPP,
explicit expressions for \emph{two new nonlinear anisotropic Hall
resistivities} $\rho_{||}^-$ \emph{and} $\rho_\perp^-$ were
derived and analyzed. The physical origin of these \emph{odd
}contributions caused by the subtle interplay between even effect
of vortex guiding and the odd Hall effect. Both new resistivities
were going to zero in the linear regimes of the vortex motion (
i.e. in the thermally activated flux-flow and ohmic flux-flow
regimes and had a bump-like current or temperature dependence in
the vicinity of highly nonlinear resistive transition from the
thermally activated flux-flow to flux-flow regimes. As the new odd
resistivities arose due to the Hall effect, their characteristic
scale was proportional to the small Hall constant as for ordinary
odd Hall effect investigated earlier in \cite{mawatari}.

In contrast to the model which  uses the uniaxial symmetric
 PPP \cite{ShSo} with the Hall effect, we consider below the more
simple modified model with asymmetric (ratchet) sawtooth washboard
pinning potential where the Hall effect is absent. It will be
shown the appearance of two step-like and two bump-like
singularities in the $\rho_{\|,\perp}^+$ and $\rho_{\|,\perp}^-$
(Hall-like) resistive responses in this model, even in the absence
of the Hall effect.

\begin{figure}
    \begin{center}
    \epsfig{file=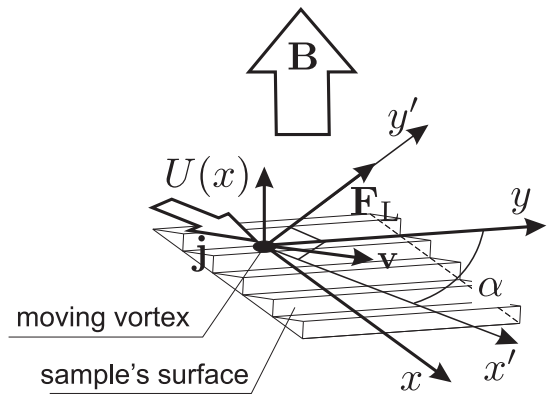,width=0.4\textwidth}
    \caption{Diagram of a superconductor in the presence
    of an external magnetic field $\mathbf{B}$. A direct transport current
    with density $\mathbf{j}$ flowing along the $x'$ direction induces a Lorentz
    force $\mathbf{F}_\textnormal{L}$ that acts along the $y'$
    direction. The superconductor is patterned
     with an asymmetric uniaxial planar pinning potential
     (asymmetric PPP) \(U(x,y)=U(x)\neq U(-x)\), whose shape is shown in the Fig.\ref{Potential}.
     The potential is periodic along the $x$-axis and translationary invariant along the $y$-axis.}
    \label{Model}
    \end{center}
\end{figure}

\begin{figure}
    \begin{center}
    \epsfig{file=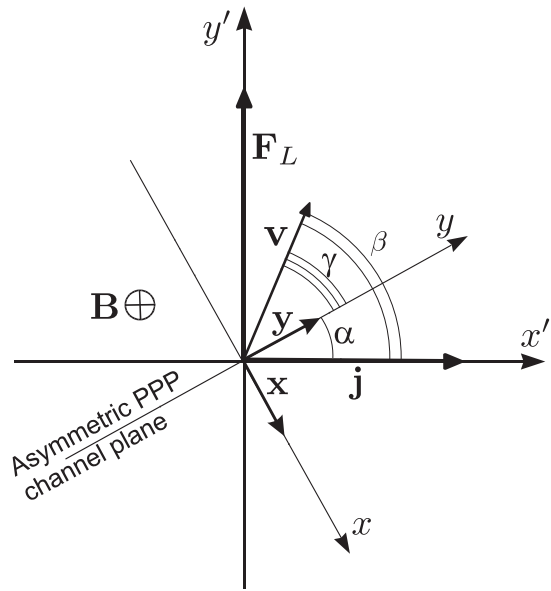,width=0.4\textwidth}
    \caption{System of coordinates $xy$ (with the unit vectors $\mathbf{x}$ and $\mathbf{y}$) associated
     with the asymmetric PPP (the unit vector $\mathbf{y}$ points along
     the pinning channels) and the system of coordinates $x'y'$ associated with
      the direction of the transport current density vector
      $\mathbf{j}$ which
         points along the $x'$ axis; $\alpha$ is the angle between the asymmetric
          PPP channels direction and the transport current density vector $\mathbf{j}$
          , $\beta$ is the angle between the average velocity vector $\mathbf{v}$ of the vortices
              and the transport current density $\mathbf{j}$; $\mathbf{F}_\textnormal{L}$ is the Lorentz force
               and $\gamma$ is the angle between the average velocity vector $\mathbf{v}$ and asymmetric PPP channel plane.}
    \label{SysCoord}
    \end{center}
\end{figure}

\begin{figure}
    \begin{center}
    \epsfig{file=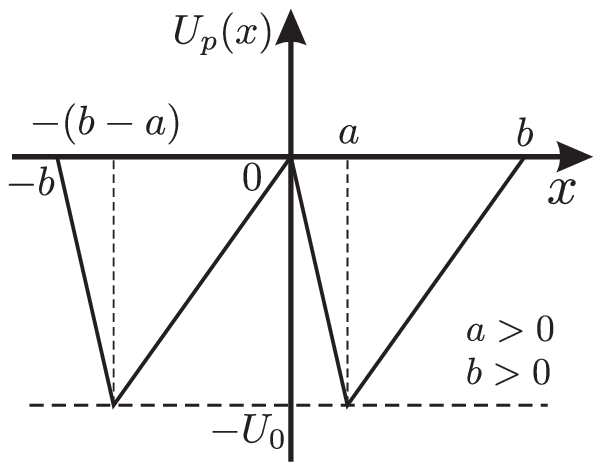,width=0.4 \textwidth}
    \caption{Asymmetric sawtooth pinning potential $U_p(x)$: $b$ is the potential period
    (width of the potential channels), $a$ is the $x$-coordinate of the minimum
    of the potential well, $U_0$ is the depth of the potential well.}
    \label{Potential}
    \end{center}
\end{figure}

The objective of this paper is to present results of a
temperature-dependent theory for the calculation of the nonlinear
magnetoresistivity tensor for asymmetric sawtooth washboard
pinning potential at arbitrary value of asymmetry parameter
$0<\varepsilon<1$ for the case of in-plane geometry of experiment.
This approach will give us the experimentally important
theoretical model which demonstrates the
$\rho^\pm_{\parallel,\perp}$ magneoresistivities for all
corresponding values of the modeling parameters and predicts an
appearance of the nonlinear magnetoresistivity $\rho_\perp^-$ at
some set of parameters $\varepsilon$ (when the \emph{Hall
coefficient} is zero) due to the asymmetry of the washboard PPP.

The organization of the  article is as follows. The section
\ref{GR}  presents those general results in the stochastic model
of anisotropic pinning which don't require specification of the
form of the pinning potential: the Fokker-Planck method in the
two-dimensional model of anisotropic pinning and the nonlinear
resistivity and conductivity tensors. In subsection
\ref{discussion} we substitute a specific sawtooth form of the
pinning potential into the general formulas of the preceding
section. It enables us to find the exact analytical solution of
our model. It will be analyzed there the behavior of the model
depending on the model's parameters: $j$, $t$, $\alpha$,
$\varepsilon$. Some formulas will be obtained for the
$\rho^\pm_{\parallel,\perp}(j,t,\alpha,\varepsilon)$. Subsection
\ref{Geffect} is dedicated to an analysis of the nonlinear guiding
effect in presence of the  PPP asymmetry, and Subsection
\ref{Rcharact} discusses the behavior of resistive responses due
to the asymmetry of pinning potential. Subsection \ref{Mrstabil}
discusses a magnetoresistivity stability with respect to small
deviations of the angle $\alpha$ from its values adopted in the
longitudinal (L) and transverse (T) geometries of experiment.
Subsection \ref{weakassym} gives a short discussion of main new
features for the case of weak asymmetry. Finally, the last section
\ref{end} represents the obtained results  and formulates the
conclusions.

\section{GENERAL RESULTS}\label{GR}
\subsection{The Fokker-Plank method\\ in the anisotropic pinning
model}\label{FPmetod}

 The Langevin equation for a vortex moving with velocity
$\mathbf{v}$ in a magnetic field $\mathbf{B}=\mathbf{n}B$
($B\equiv|\mathbf{B}|$, $\mathbf{n}=n\mathbf{z}$, $\mathbf{z}$ is
the unit vector in the $z$-direction and $n=\pm 1$) has the form
\begin{equation}
\label{F1} \eta\mathbf{v}=\mathbf{F}_{\textnormal
L}+\mathbf{F}_{p}+\mathbf{F}_{th},
\end{equation}
\ where $\mathbf{F}_{\textnormal
L}=n(\Phi_{0}/c)\mathbf{j}\times\mathbf{z}$ is the Lorentz force
($\Phi_{0}$ is the magnetic flux quantum, $c$ is the speed of
light, and $\mathbf{j}$ is the transport current density),
$\mathbf{F}_{p}=-\nabla U_{p}(x)$ is the anisotropic pinning force
($U_{p}(x)$ is the uniaxial and asymmetric ($U_p(x)\neq U_p(-x)$)
planar pinning potential),
 $\eta$ is the electronic viscosity constant. The thermal fluctuation
 force $\mathbf{F}_{th}$ is represented by a Gaussian white noise,
 whose stochastic properties are assigned by the relations
\begin {equation}
\label{F2} \langle F_{th,i}(t)\rangle=0, \, \langle F_{th,i}(t)
F_{th,j}(t') \rangle=2T\eta\delta_{ij}\delta(t-t'),
\end{equation}
where $T$ is the temperature in energy units. Employing relation
(\ref{F2}), we can reduce Eq. (\ref{F1}) to a system of
Fokker-Plank equations:
\begin{equation}
\label{vflow}
    \frac{\partial P}{\partial t}=-\nabla\cdot{\mathbf S}
\end{equation}
\begin{equation}
\label{f-p}
    \eta{\mathbf S}=({\mathbf F}_\textnormal L+{\bf F}_p)P-T\nabla P,
\end{equation}
where $P=P(\mathbf{r},t)$ is the probability density associated
with finding the vortex at the point $\mathbf r= \mathbf r(x,y)$
at the time $t$, and
\begin{equation}
\label{probab}
    \mathbf S(\mathbf r,t)\equiv P(\mathbf r,t)\mathbf v(\mathbf
    r,t),
\end{equation}
is the probability flux density of the vortex. Since the
anisotropic pinning potential is assumed to depend only on the $x$
coordinate and is assumed to be periodic [$U_p(x)=U_p(x+a),$ where
$a$ is the period], the pinning force is always directed along the
$x$ axis (with the unit anisotropy vector $\mathbf x$, see Fig.
\ref{SysCoord}) so that it has no component along the $y$ axis
[$F_{py}=-dU_p(x)/dy=0$]. Thus, Eq. (\ref{f-p}) in the stationary
case for the functions $P=P(x)$ and $\mathbf
S=(S_x,S_y)=S_x(x)\mathbf x+S_y(y)\mathbf y$ reduces to the
equations
\begin{equation}
\label{f6}
    \eta {S_x}=P(x)\left(F_{\textnormal Lx}-\frac{dU_p}{dx}
    \right)-T\frac{dP}{dx},
\end{equation}
\begin{equation}
\label{f7}
 \eta{S_y}=P(x)F_{\textnormal Ly},
\end{equation}
where $F_{\textnormal Lx}=n({\Phi_0}/{c}) j \cos \alpha$ and
$F_{\textnormal Ly}=-n({\Phi_0}/{c})j \sin\alpha$ are the $x$ and
$y$ components of the Lorentz force, respectively, and $\alpha$ is
the angle between the direction of the transport current density
$\mathbf j$ and the $y$ axis (see Figs.
\ref{Model},\ref{SysCoord}). Invoking the condition of
stationarity for Eq. (\ref{vflow}) and eliminating $S_y$ from Eq.
(\ref{f6}) and (\ref{f7}) we obtain
\begin{equation}
\label{f8}
    T \frac{dP}{dx}+\left(\frac{dU_p}{dx}-F_{\textnormal Lx}\right)P=-S_x\eta
\end{equation}
From the mathematical point of view, Eq. (\ref{f8}) is the
Fokker-Planck equation of the one-dimensional vortex dynamics. The
solution of Eq. (\ref{f8}) for periodic boundary conditions
$P(0)=P(a)$ and one-dimensional periodic pinning potential of
general form is
\begin{equation}
\label{f9}
    P(x)=\frac{\eta S_x}T \frac{f(a)f(x)}{f(a)-f(0)} \int_x^{x+a}
    \frac{d\xi}{f(\xi)},
\end{equation}
where $f(\xi)=exp([F\xi-U_p(\xi)]/T)$.

Using the definition of the mean vortex velocity
\begin{equation}
\label{f10}
    \langle \mathbf v \rangle = \int_0^a \mathbf S(x)dx / \int_0^a
    P(x)dx,
\end{equation}
we obtain an expression for the $x$ and $y$ components of the
vortex mean velocity
\begin{equation}
\label{f11}
    \langle \upsilon_x \rangle = \int_0^a S_x(x) dx / A = S_x a / A =
    F_{\textnormal Lx}\nu(F_{\textnormal Lx})/\eta,
\end{equation}
\begin{equation}
\label{f12}
    \langle \upsilon_y \rangle = \int_0^a S_y(x) dx / A =
    F_{\textnormal Ly}/\eta,
\end{equation}
where $A=\int_0^a P(x)dx$ and
\begin{multline}
\label{f13}
    \frac{1}{\nu(F_{\textnormal Lx})} \equiv \frac{F_{\textnormal Lx}}{Ta(1-\exp(-F_{\textnormal Lx}a/T))}\int_0^a dx \\
    \times \int_0^a dx' \exp\left(-\frac{F_{\textnormal Lx}x}T\right) \\ \times
    \exp\left(\frac{U_p(x+x')-U_p(x')}{T}\right).
\end{multline}
The dimensionless function $\nu(F_{\textnormal Lx})$ in the limit
$F_{\textnormal Lx} \to 0$ coincides with the analogous quantity
introduced in \cite{mawatari}. It has the physical meaning of the
probability of the vortex overcoming the potential barrier, the
characteristic value of which we denote as $U_0$. This can be seen
by considering the limiting cases of high ($T \gg U_0$) and low
($T \ll U_0$) temperatures. In the case of high temperatures we
have $\nu \approx 1$, and expression (\ref{f13}) corresponds to
the flux-flow regime (FF regime). Indeed, in this case the
influence of pinning can be neglected. In the case of low
temperatures $\nu$ is a function of the transport current. For
strong currents ($Fa \gg U_0$) the potential barrier disappears,
$\nu \approx 1$, and the FF regime is realized. For weak currents
($Fa \ll U_0$) we have $\nu \sim \exp(-U_0/T)$, which corresponds
to the thermally activated flux-flow regime (TAFF regime). The
transition from the TAFF regime to the FF regime is associated
with a lowering of the potential barrier with growth of the
current.


\subsection{The nonlinear conductivity \\ and resistivity
tensors}\label{resist}

The average electric field in the $xy$ coordinate system induced
by the moving vortices is given by
\begin {equation}
\label{f14} \mathbf{E}=(1/c)\mathbf{B}\times \langle\mathbf{v}
\rangle=n(B/c)(-\langle\upsilon_{y}\rangle\mathbf{x}+\langle\upsilon_{x}\rangle\mathbf{y}).
\end {equation}

Taking Eqs. (\ref{f11}), (\ref{f12}) and (\ref{f14}) we obtain the
dimensionless magnetoresistivity tensor $\hat{\rho}$ (having
components measured in units of the flux-flow resistivity
$\rho_{f}$) for the nonlinear law
$\mathbf{E}=\hat{\rho}(j)\mathbf{j}$
\begin{equation}
\label{f15}
\begin{array}
{l}\hat{\rho}=
\left(
    \begin {array}{cc} \rho_{xx}& \rho_{xy}\\
    \rho_{yx}& \rho_{yy}
    \end {array}
\right )= \left(
    \begin {array}{cc}
    1 & 0\\
    0 & \nu(f)
    \end{array}
\right ),
\end{array}
\end{equation}
where the dimensionless components of the electric field are
measured in units of $E_0=BU_0/ca\eta$, and of the current, in
units of $j_0=cU_0/\Phi_0b$, and
\begin{multline}
\label{f21}
    f=-\frac{F_{\textnormal Lx}b}{U_0}=m\cdot n\cdot j_y = \\
    =m\cdot n\cdot j \cdot \cos\alpha=p\cdot j\cdot \cos\alpha,
\end{multline}
where $m=\pm 1$ determines the transport current reversal (
$j=m|\mathbf j|$), $n=\pm 1$ determines the magnetic field
direction reversal ($\mathbf B=n|\mathbf B|$), $p\equiv m\cdot n$
is the combination for simplification of the current and magnetic
field directions reversal and $\alpha$ is the angle between the
current direction and asymmetric PPP channels.

The conductivity tensor $\hat{\sigma}$ (the components of which
are measured in units of $1/\rho_{f}$), which is inverse of the
tensor $\hat{\rho}$, has the form
\begin {equation}
\label{f16} \begin{array}{l}\hat{\sigma}=\left(\begin {array}{cc} \sigma_{xx}& \sigma_{xy}\\
\sigma_{yx}& \sigma_{yy} \end {array}\right )=\left(\begin
{array}{cc} 1 & 0 \\
0 & \nu^{-1}(f) \end {array}\right ).
\end{array}
\end {equation}

From Eqs. (\ref{f15}) and (\ref{f16}) we see, that off-diagonal
components of the $\hat{\rho}$ and $\hat{\sigma}$ tensors are
zero, and the nonlinear components of the $\hat{\rho}$-tensor and
$\hat{\sigma}$-tensor are functions of the external force value
$f$ through the external current density $\mathbf j$, the
temperature $T$, and the angle $\alpha$.

The experimentally measurable resistive responses refer to a
coordinate system tied to the current (see Fig. \ref{SysCoord}).
The longitudinal and transverse (with  respect to the current
direction) components of the electric field, $E_{\parallel}$ and
$E_{\perp}$, are related to $ E_{x}$ and $E_{y}$ by the simple
expressions
\begin{equation}
\label{f17} \left \{
\begin{array}{lll}
 E_{\parallel}&=&E_{x}\sin\alpha+E_{y} \cos \alpha,\\
 \\
 E_{\perp} & = & -E_{x}\cos\alpha+E_{y}\sin\alpha.\\
\end{array}
\right.
\end{equation}

Then according to Eqs. (\ref{f15}) and (\ref{f17}), the
expressions for the experimentally observable longitudinal and
transverse (with respect to the $\mathbf{j}$-direction )
magnetoresistivities $\rho_{\parallel}\equiv E_{\parallel}/j $ and
$\rho_{\perp}\equiv E_{\perp}/j$ have the form:
\begin{equation}
\label{f18} \left \{
\begin{array}{lll}
 \rho_{\parallel}&=&\sin^2\alpha+\nu(f)\cos^2\alpha,\\
 \\
 \rho_{\perp} & = & (\nu(f)-1)\cos\alpha \sin\alpha.\\
\end{array}
\right.
\end{equation}

We introduce the L and T geometries in which $\mathbf
j\parallel\mathbf x$ and $\mathbf j\perp\mathbf x$, respectively.
From Eq. ($\ref{f18}$) follows, that in the L geometry vortex
motion takes place along the pinning channels (the guiding
effect), and in the T geometry - transverse to the pinning
channels direction (the slipping effect). In the L geometry the
crossover( or critical at $T>0$) current is equal to zero since
the FF regime is realized for guided vortex motion along pinning
channels direction . In the T geometry, i.e., for the vortex
motion transverse to the pinning channels, a pronounced nonlinear
regime is realized for $T\ll U_0$, in the range
$j_{cr1}<j<j_{cr2}$ (we denote it by the $cr$ (crossover) index).
In our case, when the temperature $T>0$, strictly speaking there
is no exact value of the crossover currents $j_{cr1}$ and
$j_{cr2}$, as at any relatively low temperature vortices can move
transverse to the pinning channels. However, when $T\ll U_0$, the
existence of the crossover current might make sense which
separates the guiding region of motion of the vortices, where they
move only along the pinning channels, and the slipping region when
vortices also slip over pinning barriers.

It is evident that the presence of different crossover currents
for mutually opposite directions along the vector $\mathbf x$ is a
direct consequence of an asymmetric pinning potential.
\begin{figure}
    \begin{center}
    \epsfig{file=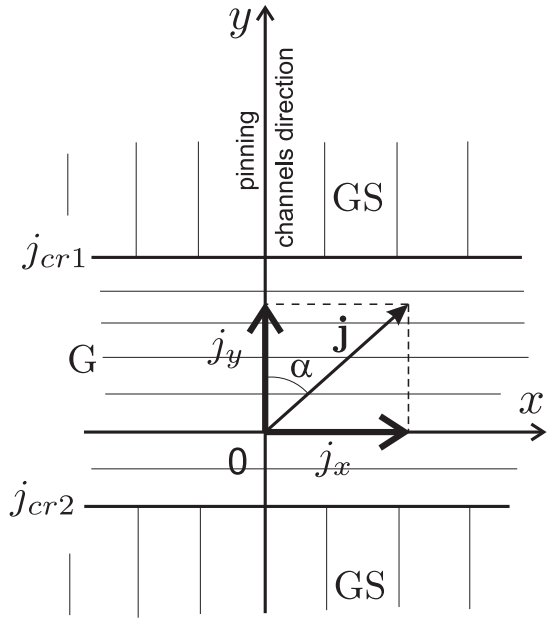,width=0.4 \textwidth}
    \caption{Diagram of dynamic states of the vortex system in the
   $xy$ plane; G is the region of motion of the vortices along the
    pinning channels (the guiding effect), GS is the region of motion
    of the vortices along and transverse to the pinning channels
    (the guiding and slipping effects together); $j_{cr1}$ and $j_{cr2}$
    are the crossover currents for mutually opposite directions along the vector $\mathbf y$
         corresponding to a transition from the region G to
          the region GS when the $j_y$ is increasing.}
    \label{diagram}
    \end{center}
\end{figure}
Let us consider a diagram of the dynamical states of the vortex
system in the $(j_x,j_y)$ plane (Fig. \ref{diagram}). For
arbitrary angle $\alpha$ the tip of the vector $\mathbf j$ can lie
in two different regions which are different in their physical
meanings. If $j_y<j_{cr1}$ or $j_y<j_{cr2}$ the guided vortex
motion takes place (the guiding region). For $j_y>j_{cr1}$ or
$j_y<j_{cr2}$ the guided motion along the pinning channels is
joined by motion transverse to the pinning channels (the slipping
region). It is clear, that if we apply an alternating current
$j_y^{AC}$ along the $y$ axis such that the amplitude of a current
satisfies to the relation $j_{cr1}<j_y^{AC}<j_{cr2}$, it will lead
to a motion of vortices along the $x$ axis as they can overcome a
pinning potential only in one direction. It is the occurrence of
the ratchet effect.


\section{VORTEX PINNING on ASYMMETRIC PPP and ANALYSIS of NONLINEAR
REGIMES.}\label{APPP}

\subsection{Pinning potential and ${\nu}$-function behavior
.}\label{discussion}

The nonlinear properties of the resistivity tensor $\hat \rho$ ,
as can be seen from formula (\ref{f15}), are completely determined
by the behavior of the function $\nu$, which has the physical
sense of the probability of a vortex overcoming the potential
barriers created by the channels of asymmetric pinning potential.
In turn, the function $\nu$, according to formula (\ref{f13}),
depends on the form of the pinning potential. Above we considered
the simplest case of the asymmetric PPP that has sawtooth-like
form (see Fig. \ref{Potential})
\begin{equation}
\label{f19} U_p(x)= \left\lbrace
    \begin{array}{lll}- F_{p1}x, & 0\leqslant x < a\\
               - F_{p2}(x-b), & a < x \leqslant b,
    \end{array}
    \right.
\end{equation}
where $F_{p1}=U_0/a$ and $F_{p2}=-U_0/(b-a)$ are the pinning
forces in the different directions of the $x$ - axis and $U_0$ is
the depth of a potential well, $b$ is the period of  the
asymmetric PPP ($b\geqslant a$), $\varepsilon=a/b$ is the
parameter characterizing the asymmetry of the pinning potential
($0<\varepsilon<1$ and $\varepsilon=1/2$ corresponds to the
symmetric well).

Substituting the potential (\ref{f19}) into formula (\ref{f13})
for the probability function $\nu$ gives the following expression:
\begin{multline}
\label{f20}
    \nu(f,t,\varepsilon)=\left(
        (\exp(f/t)-1)(1-2f\varepsilon+2f+f^2\varepsilon^2 \right.
        \\ \left. -2f^2\varepsilon+f^2)(f \varepsilon - 1)^2 \right) /
        \left(f(-1 \right. \\
        \left. +6\exp(f/t)f^2\varepsilon^2+2\varepsilon+2f^2\varepsilon \right. \\
        \left. -5f\varepsilon^2+t\exp((f\varepsilon-f-1)/t) - t \right. \\
        \left. +t\exp((f\varepsilon-1)/t)-f-\varepsilon^2f^3+2\varepsilon^3f^3 \right. \\
        \left. +5f\varepsilon+f\exp(f/t)+4\varepsilon^3f^2 \right. \\
        \left. -2f^2\varepsilon\exp(f/t)+ f^3\varepsilon^2\exp(f/t) \right. \\
        \left. +f^3\varepsilon^4\exp(f/t)-\varepsilon^4f^3+\exp(f/t) \right. \\
        \left. -t\exp(f/t)-4f^2\varepsilon^3\exp(f/t) \right. \\
        \left. -2f^3\varepsilon^3\exp(f/t)-6f^2\varepsilon^2-2\varepsilon\exp(f/t) \right. \\
        \left. +5f\varepsilon^2\exp(f/t)-5f\varepsilon\exp(f/t) \right),
\end{multline}
where $f=-F_{Lx}b/U_0$ is the dimensionless external force which
gives ratio of this force to the average pinning force $U_0/b$,
$t=T/U_0$ is the dimensionless temperature which gives the ratio
of the energy of the thermal fluctuations of the vortices to the
depth of the potential wells $U_0$. In our case the dimensionless
external force $f$ also coincides with the dimensionless transport
current $j_y$, which is given by formula (\ref{f21}).

At function evaluation (\ref{f20}) we assumed, that the asymmetry
parameter $\varepsilon$ changes from zero value, that corresponds
to shift of the pinning potential minimum to the left, up to unity
value that corresponds to shift of potential to the right and that
naturally leads to as much as possible asymmetry of the pinning
potential in appropriate direction.

Let us consider in turn the dependence of the probability function
$\nu(f,t,\varepsilon)$ on each of the quantities $f$, $t$ and
$\varepsilon$ for the remaining quantities held fixed (denoted by
the subscript "0").

\begin{figure}[t]
    \begin{center}
    \epsfig{file=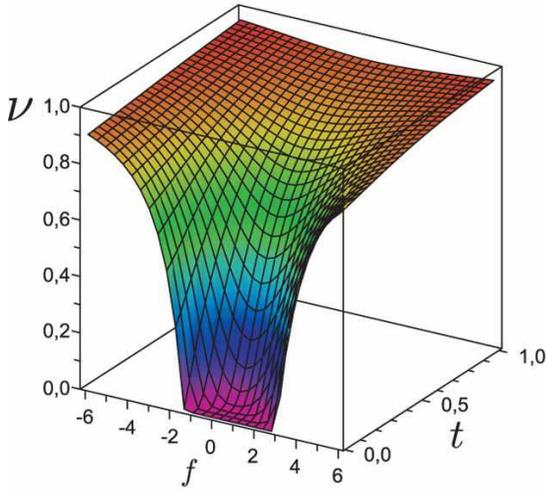,width=0.4 \textwidth}
    \caption{The dependence $\nu(f,t)$ for fixed value of the asymmetry parameter $\varepsilon_0=1/3$.}
    \label{V3d}
    \end{center}
\end{figure}

The dependence $\nu(f,t)=\nu(f,t,\varepsilon_0)$ (see Fig.
\ref{V3d}) characterizing $\nu$ as a function of the external
force acting on a vortex at constant asymmetry parameter. The
influence of the external force $f$ acting on the vortices is that
it lowers the height of the potential barrier for vortices
localized along the channels of the asymmetric PPP and,
consequently, increases the probability of escape from them.
Raising the temperature also increases the probability that a
vortex will escape from a potential well through an increase in
the energy of the thermal fluctuations of the vortices. Thus, the
pinning potential, leading as $f,t \to 0$ to localization of
vortices, can be suppressed both by an external force and by an
increase in the temperature. From Eq. (\ref{f20}) follows that
\begin{equation}
\label{f22}
    \lim_{t \to \infty}\nu(f,t,\varepsilon)=\lim_{f \to \infty}\nu(f,t,\varepsilon)=1.
\end{equation}

\begin{figure}[t]
    \begin{center}
    \epsfig{file=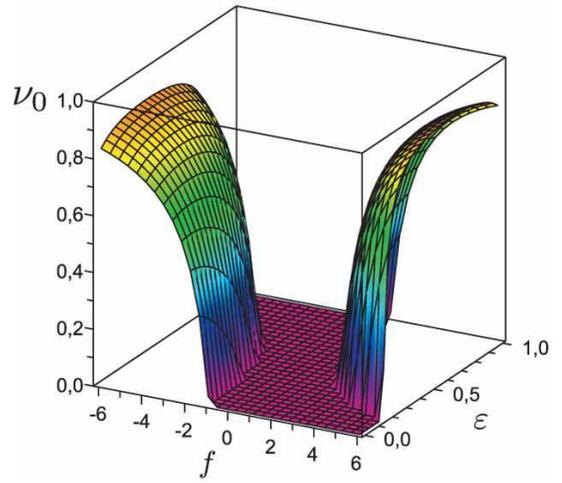,width=0.4 \textwidth}
    \caption{The dependence $\nu_0(f,\varepsilon)=\lim_{t \to 0}\nu(f,t,\varepsilon)$ at
    zero temperature limit.}
    \label{V3d_by_zero_t}
    \end{center}
\end{figure}

The function
\begin{equation*} \nu_0(f,\varepsilon)=\lim_{t \to
0}\nu(f,t,\varepsilon)
\end{equation*}
is equal to
\begin{equation}
\label{f23} \nu_0(f,\varepsilon)= \left\lbrace
    \begin{array}{cl} 0 & \frac{1}{\varepsilon-1}<f<\frac{1}{\varepsilon}\\
                \frac{(f\varepsilon-f-1)(f\varepsilon-1)}{f(f\varepsilon^2-
                f\varepsilon-2\varepsilon+1)}, & (f<\frac1{\varepsilon-1})\cup (f>\frac1{\varepsilon}),
    \end{array}
    \right.
\end{equation}
corresponds to the zero-temperature limit (see Fig.
\ref{V3d_by_zero_t}).

From Eqs. (\ref{f23}) and (\ref{f21}) follows, that the crossover
transport current and, respectively, crossover external force (in
the dimensionless units) in both directions is
$f_{cr1}=j_{cr1}=1/(\varepsilon-1)$ and
$f_{cr2}=j_{cr2}=1/\varepsilon$ (see also the diagram of dynamic
states in Fig. \ref{diagram}). If $0<\varepsilon<1/2$, then
$|j_{cr1}|<|j_{cr2}|$, and for $1/2<\varepsilon<1$ we have
$|j_{cr1}|>|j_{cr2}|$.

In the zero-temperature limit, for $|j_y|<|j_{cr1}|,|j_{cr2}|$ the
vortices are trapped in the potential wells of the pinning
channels and they cannot move across pinning barriers, while for
$|j_y|>|j_{cr1}|,|j_{cr2}|$ the potential barrier disappears and
the vortices begin to move in the one or both directions.

Let's note also, that value $\varepsilon=1/3$ which we use in some
figures of this article, corresponds to a case when
$j_{cr1}=2j_{cr2}$.

It is easy to understand the influence of the temperature on the
qualitative form of $\nu(j_y,\varepsilon)$. Specifically at low
temperatures ($T\ll U_0$) a nonlinear transition takes place from
the TAFF regime of vortex motion perpendicular to the pinning
channels to the FF regime with growth of the external force,
wherein the function $\nu(j_y,\varepsilon)$ has a characteristic
nonlinear shape (see Fig. $\ref{V3d}$). At high temperatures ($T
\gg U_0$) the FF regime is realized over the entire range of
variation of the external current. At nonzero temperature the
$j_{cr1}$ and $j_{cr2}$ disappear because $\nu(j_y,\varepsilon)$
is not vanishing at any value of the parameters $j_y,\varepsilon$.
Increase of the temperature leads to smoothing of the function
$\nu(j_y,\varepsilon)$ and in the limit $t\to \infty$ it simply
degenerates in plane $\nu(j_y,\varepsilon)=1$ that corresponds to
the free motion of the vortices.

In the limit $f \to 0$ we have:
\begin{equation}
\label{f24}
    \nu_0(t,\varepsilon)=\frac{\exp(1/t)}{t^2(\exp(1/t)-1)^2}.
\end{equation}
Note, that the value of $\nu_0(t,\varepsilon)$ does not depend on
the parameter of asymmetry of a pinning potential. Physically it
means that effects related to asymmetry of a pinning potential are
relevant when the vortices are moving (they are moving always
except the case when $T=0$ and $|j|<|j_{cr1}|,|j_{cr2}|$).

As follows from Eqs. (\ref{f21}),(\ref{f18}) and (\ref{f20}) the
dynamics of a vortex system depends substantially on the direction
of the current flow and the magnetic field reversal. According to
Eq. (\ref{f21}), an external transport current density $\mathbf j$
or direction of the external magnetic field $\mathbf n$ equally
cause the reversal of the Lorentz force $F_\textnormal L$ which
change the magnitude of the $\nu(f,t,\varepsilon)$ function due to
the inversion of \emph{f}. In order to consider only
\emph{p}-independent magnitudes of the $\rho_{\parallel}$ and
 $\rho_{\perp}$ resistivities in Eqs. (19) we should introduce
the even(+) and odd(-) magnetoresistivities with respect to
\emph{p}-inversion ($(\rho(p)^{\pm}\equiv
(\rho(p)\pm\rho(-p))/2)$). From this point of view follows that we
should present the function $\nu(f)$ as a sum of the even (+)and
odd (-) parts with respect to inversion of the moving force:
\begin{equation}
\label{f25}
    \nu(p)=\nu^+(p) + \nu^-(p),
\end{equation}
\begin{equation}
\label{f26}
    \nu^{\pm}(p)=\frac{\nu(pf,t,\varepsilon)\pm\nu(-pf,t,\varepsilon)}{2},
\end{equation}
where $\nu^{\pm}$ are even and odd parts of the $\nu$ function
respectively.

\begin{figure}[t]
    \begin{center}
    \epsfig{file=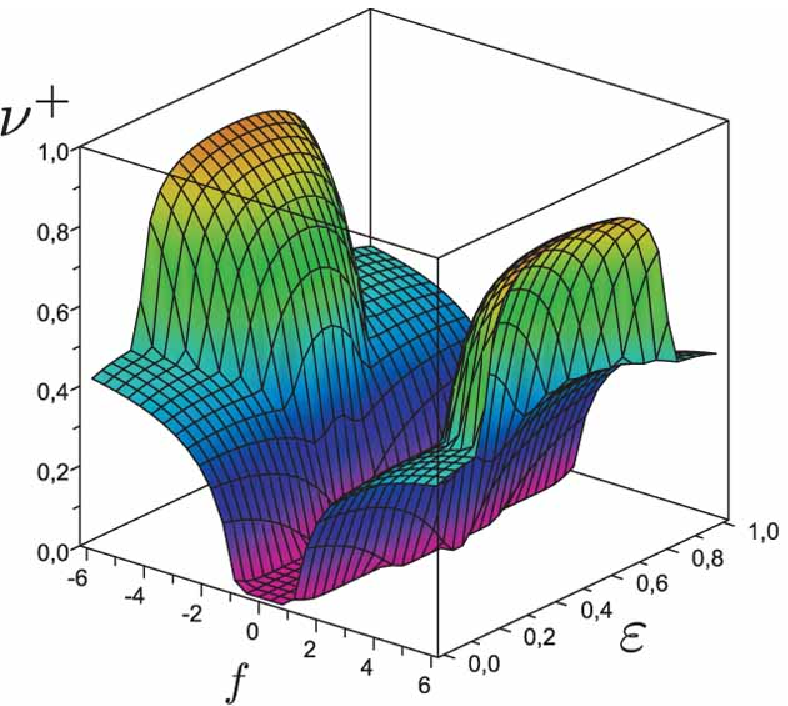,width=0.4 \textwidth}
    \caption{The dependence $\nu^+(f,\varepsilon)$ for fixed value of the temperature $t_0=0.05$.}
    \label{Vp3d}
    \end{center}
\end{figure}

\begin{figure}[t]
    \begin{center}
    \epsfig{file=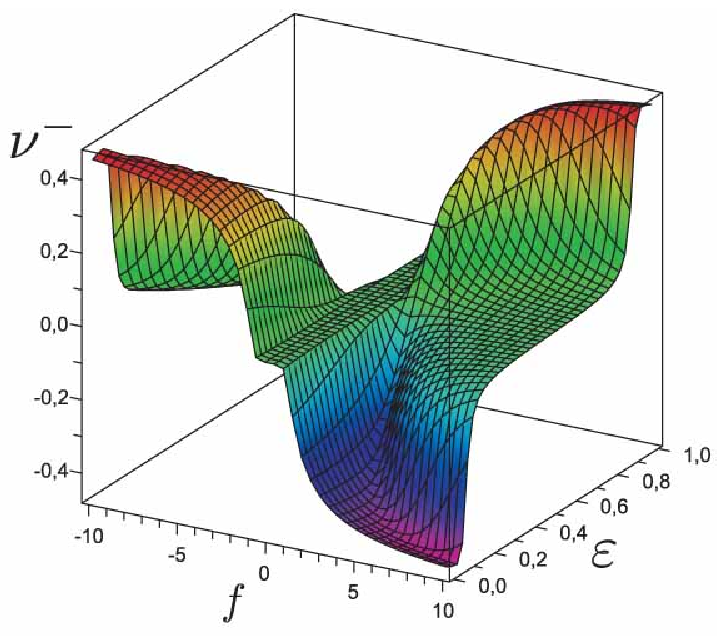,width=0.44 \textwidth}
    \caption{The dependence $\nu^-(f,\varepsilon)$ for fixed value of the temperature $t_0=0.05$.}
    \label{Vme3d}
    \end{center}
\end{figure}

As we can make sure, the dependence $\nu^+(f,t)$ is closely
similar to the $\nu(f,t)$. The qualitative behavior and the limits
of the component $\nu^+(f,t)$ as $f,t \to 0,\infty$ coincide with
the corresponding limits of $\nu(f,t)$ as it is follows from Eq.
(\ref{f22}) and Eq. (\ref{f26}).

The dependences of the
$\nu^\pm(f,\varepsilon)=\nu^\pm(f,t_0,\varepsilon)$ as a function
of the external motive force and asymmetry parameter at a constant
temperature are shown in Figs. \ref{Vp3d} and \ref{Vme3d}. At
comparatively low temperatures ($t \ll 1$) the asymmetry parameter
$\varepsilon$ influences to shape of the $\nu^+$ function (see
Fig. \ref{Vp3d}). It is appears as step-like dependence of the
$\nu^+(f)$ at $\varepsilon \neq 1/2,0,1$. The origin of this steps
corresponds to critical currents density $j_{cr1,2}$. Such, that
if $\varepsilon = 1/2$ than we have one step with origin at
$f_{cr1}=f_{cr2}=2$ and function will tend up to unity if
$f\rightarrow 0$. In case of $\varepsilon = 0$ or $\varepsilon=1$
corresponding values of the critical current density are equal to
$f_{cr1}=-1, f_{cr2}=+\infty$ or $f_{cr1}=-\infty, f_{cr2}=1$
respectively, and we should see one step again and the value of
the $\nu^+$ function does not exceed $1/2$ . In the other case we
have two step in the $\nu^+(f)$ dependence as consequence of the
asymmetry of the pinning potential. It is worth noticing, that
function under consideration is symmetrical relatively to planes
$f=0$ and $\varepsilon=1/2$.

\begin{figure}[t]
    \begin{center}
    \epsfig{file=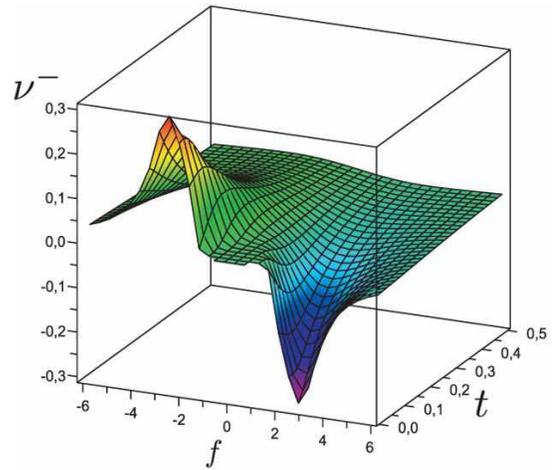,width=0.4 \textwidth}
    \caption{The dependence $\nu^-(f,t)$ for fixed value of the asymmetry parameter $\varepsilon_0=1/3$.}
    \label{Vm3d}
    \end{center}
\end{figure}

\begin{figure}[t]
    \begin{center}
    \epsfig{file=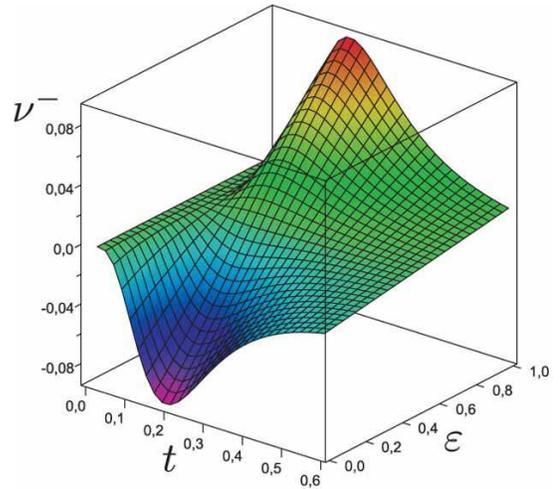,width=0.4 \textwidth}
    \caption{The dependence $\nu^-(t,\varepsilon)$ for fixed value of the external motive force $f_0=0.7$.}
    \label{Vmet3d}
    \end{center}
\end{figure}

The component $\nu^-(f,t)$ tends to zero in the linear regimes (as
$f,t \to 0,\infty$) and is nonzero in the region of nonlinearity
of $\nu$ (see Fig. \ref{Vm3d}). It means that $\nu^-(f,t)$ can be
suppressed with increasing of the external motive force or
temperature. When the value of temperature grows, function
$\nu^-(f,t)$ tends to zero since the temperature fluctuations at
$\emph{t} \gg 1$ in a superconductor suppress the influence of the
pinning potential and the vortex, thus, can move freely in any
direction (flux-flow regime is realized). At small temperature,
when the contribution of pinning potential is comparable or more
than the contribution of temperature fluctuations ($t\ll 1$), the
$\nu^-$ behavior will explain by interplay of the external motive
force $f$ and asymmetry pinning potential (which causes to
appearance the critical currents $j_{cr1,cr2}$). If $f\ll
f_{cr1},f_{cr2}$ and the temperature value is close to zero, then
$\nu^-(f,t)$ also is close to zero because the vortex cannot move
across pinning channels in any direction. When $f_{cr1}<f<f_{cr2}$
(or vice versa $f_{cr2}<f<f_{cr1}$ that depends on the value of
the asymmetry parameter $\varepsilon$) than the thermally
activated flux-flow regime is realized and the curve $\nu^-(f,t)$
accepts a bump-like form (see Fig. \ref{Vmet3d}). In this case the
ratchet effect occurs and vortices start to move in a direction
that associated with the minimal pinning force. At strong external
motive force, when $f\gg f_{cr1},f_{cr2}$, the function $\nu^-(f)$
also is vanishing because external current suppresses influence of
the pinning potential on the vortex ( flux-flow regime arises) as
it follows from the general properties of the $\nu(f)$ function
and the vortices can freely move in any direction across the
pinning channels. The width of the base of a bump-like curve is
associated with asymmetry parameter $\varepsilon$ and also can be
simply presented as $2\Delta
f_{cr}=2||f_{cr1}|-|f_{cr2}||=|(|\varepsilon|-|\varepsilon-1|)/((\varepsilon-1)\varepsilon)|$.
It is means when $\varepsilon\rightarrow 0,1$ the width of the
peak will tend to infinity and in Fig. \ref{Vme3d} we seeing a
change the bump-like to step-like dependence of $\nu^-(f)$. The
maximum of the $\nu^-(f)$ function with respect to the external
motive force $f$ (see Figs. \ref{Vme3d}, \ref{Vm3d}) corresponds
to the maximum pinning force of both the $f_{cr1}$ and $f_{cr2}$,
respectively.

Appearance of the $\nu^-(f,t)$ is a direct consequence of
asymmetry of the pinning potential. When the asymmetry parameter
$\varepsilon$ is not equal to $1/2$, then the same absolute value
of a motive force enclosed in mutually opposite directions leads
to different values of the function $\nu$ that leads to occurrence
of an odd component $\nu^-$.

When the pinning potential is symmetric ($\varepsilon=1/2$), the
function $\nu^-$ is equal to zero (see Figs. \ref{Vme3d},
\ref{Vmet3d}).

Also, for $\varepsilon=1/2$ we regain the results of Ref.
\cite{UsSh}:
\begin{multline}
\label{f27}
    \nu(f,t,1/2)=(u^2-1)^2/(u^2(u^2-1) \\
        2ut(\cosh(u/t)-\cosh(1/t))/\sinh(u/t)),
\end{multline}
where $u=2f$.

In Ref. \cite{ShSo} the influence of the Hall effect on occurrence
of the $\nu^-$ in presence of symmetric pinning potential has been
discussed. It has been shown that if the Hall constant is distinct
from zero then $\nu^-$ is distinct from zero too, and vice versa.
Now we see that the asymmetric PPP leads to occurrence of an odd
component $\nu^-(f)$ if we neglect the Hall effect.

The dependence $\nu^-(f,t)=\nu^-(f,t,\varepsilon_0)$ and
$\nu^-(t,\varepsilon)=\nu^-(f_0,t,\varepsilon)$ explain behavior
of $\nu^-$ relative to the temperature, external motive force and
asymmetry parameter $\varepsilon$ for the fixed values of the
asymmetry parameter (Fig. \ref{Vm3d}) and the external motive
force (Fig. \ref{Vmet3d}). From Fig. \ref{Vm3d} follows that
$\nu^-$ function is tend to maximum if temperature tend to zero,
but in Fig. \ref{Vmet3d} we can observe extremum by temperature.
It is happens because temperature's raising firstly lead to
activation of the vortices overcoming across channel's walls of
the pinning potential and suppres the $\nu^-$ function when $t \gg
1$.

Now we will obtain expressions from formulas (\ref{f18}),
(\ref{f20}), (\ref{f21}) and (\ref{f26}) for the experimentally
observed longitudinal and transverse resistivities (relative to
the current direction) with the asymmetric pinning potential taken
into account. We separate out their even and odd components
relative to the current direction:

\begin{equation}
\label{f28}
 \rho_{\parallel}^+ = \nu^+\cos^2\alpha + \sin^2\alpha,
\end{equation}
\begin{equation}
\label{f29}
 \rho_{\perp}^+  = (\nu^+ - 1)\sin(2\alpha)/2,
\end{equation}
\begin{equation}
\label{f30}
 \rho_{\perp}^-  = \nu^-\sin(2\alpha)/2,
\end{equation}
\begin{equation}
\label{f31}
 \rho_{\parallel}^-  =  \nu^- \cos^2\alpha,
\end{equation}
where $\nu^\pm$ are the above-defined even and odd components
relative to the current direction of the function
$\nu(f,t,\varepsilon)$. In formulas (\ref{f28}), (\ref{f29}) the
nonlinear and linear terms separate out in a natural way. The
physical reason for the appearance of linear terms is that in the
model under consideration for $\alpha=\pi/2$ there is always a
flux-flow regime of vortex motion along the pinning channels.

\subsection{The peculiarities of nonlinear guiding effect in
presence of the PPP asymmetry.}\label{Geffect}

As is well known \cite{NW}, the specifics of anisotropic pinning
consist in the noncoincidence of the directions of the external
motive force acting on the vortex, and its velocity.

From Fig. (\ref{SysCoord}) and from Eqs. (\ref{f14}) and
(\ref{f18}) follows formula
\begin{multline}
\label{f32}
    \beta=\beta(j,t,\varepsilon,\alpha)=\arccot(\rho_{\perp}/\rho_{\parallel})=\\
        \arccot((1-\nu(j\cos\alpha,t,\varepsilon))/(\tan\alpha \\
        +\nu(j\cos\alpha,t,\varepsilon)\cot\alpha)),
\end{multline}
which is used to describe the guiding effect, where $\beta$ is the
angle between the average vortex velocity vector $\mathbf v$ and
the current density vector $\mathbf j$ (see Fig. \ref{SysCoord}).
The guiding effect is expressed that much more strongly, the
larger is the difference in directions of $\mathbf
F_\textnormal{L}$ and $\mathbf v$, i.e., the smaller is the angle
$\beta$. If $\beta=\alpha$, it means a full guided motion of
vortices exists (also named as G-regime) when all vortices move
parallel to the pinning channels and, on the contrary, if
$\beta=\pi/2$, the free vortices motion exists (in the FF regime).

The guiding effect conception is very important above all as
illustration of directed vortices motion. We will be able to
analyze the $\beta(j,t,\varepsilon,\alpha)$ by means of divide
$\beta$ into even and odd part, like $\nu$ and $\rho$ dependences.
The even $\beta^+$ part concerned with directed vortices motion
along pinning channels as it were discussed in detail in
\cite{ShSo}. Unfortunately, but the odd $\beta^-$ part hasn't
pictorial view, like $\beta^+$, and quantitative experimental
measurement of the $\beta^-$ with difficulty.

Now, we had to underline that guiding of the vortices in the
pinning channels is the necessary condition for appearance ratchet
effects's, but effect's magnitude entirely depends from
distinction between probability of the vortices overcoming over
pinning potential walls in both directions as it's follows from
subsection \ref{APPP}.

\subsection{The resistive responses due to asymmetry of the pinning
potential.}\label{Rcharact}

In this subsection we consider peculiarities of the resistive
characteristics in the investigated model due to the asymmetry of
the pinning potential. Experimentally, two types of measurements
of the observed resistive characteristics are possible in a
prescribed geometry defined by a fixed value of the angle
$\alpha$: CVC measurements and resistive measurements, which
investigate the dependence of the observed resistivities on the
current density at a fixed temperature
$\rho^\pm_{\parallel,\perp}(j,t_0)$ and on the temperature for
fixed current density $\rho^\pm_{\parallel,\perp}(j_0,t)$. The
form of these dependences is governed by a geometrical factor ---
the angle $\alpha$ between the directions of the current density
vector $\mathbf j$ and the PPP channels. There are two different
forms of the dependence of $\rho^\pm_{\parallel,\perp}$ on the
angle $\alpha$ (see formulas (\ref{f28}) - (\ref{f31})). The first
of these is the "tensor" dependence, also present in the linear
regimes (TAFF, FF and strong FF regimes), which is external to the
function $\nu$. The second is through the dependence of the
function $\nu$ on its argument $f=j_y=|\mathbf j|\cos\alpha$,
which in the region of the transition from the thermally activated
flux-flow to the flux-flow regime is substantially nonlinear  (see
Eq. (\ref{f20})).

First recall that in the absence of an asymmetry of the pinning
potential ($\varepsilon=1/2$) there exist only even resistivities
$\rho^+_{\parallel,\perp}$ in the magnetic field, whereas the odd
resistivities $\rho^-_{\parallel,\perp}$ are zero (see formulas
(\ref{f28}) - (\ref{f31})). The presence of $\varepsilon\neq1/2$
leads to the appearance of the odd component $\nu^-$, which has a
maximum in the region of the nonlinear transition from the TAFF to
the FF regime and is essentially equal to zero outside of this
transitional region (see Figs. \ref{Vme3d}, \ref{Vm3d}).

Let us analyze the resistive dependences
$\rho^\pm_{\parallel,\perp}(j)$ and
$\rho^\pm_{\parallel,\perp}(t)$ with allowance for the asymmetric
pinning potential. The nature of the behavior of the current and
temperature dependence of $\rho^\pm_{\parallel,\perp}$ is
completely determined by the behavior of the dependences
$\nu^\pm(j)$ and $\nu^\pm(t)$. As follows from formulas
(\ref{f28}) - (\ref{f31}), the even resistivities
$\rho^+_{\parallel,\perp}$ depend only on the even function
$\nu^+$ and similarly, $\rho^-_{\parallel,\perp}$ depend only on
the odd function $\nu^-$.

\begin{figure}[t]
    \begin{center}
    \epsfig{file=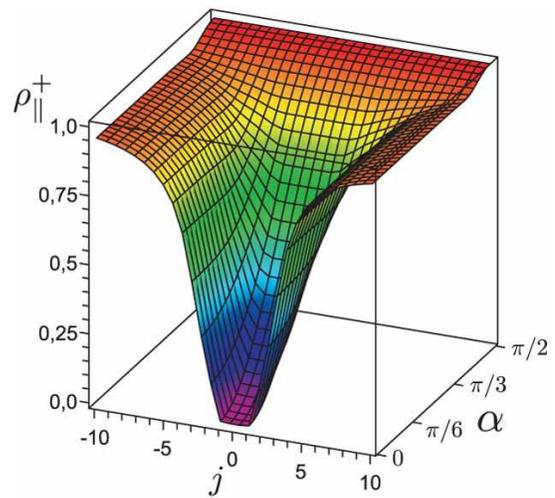,width=0.4 \textwidth}
    \caption{The dependence $\rho^+_\parallel(j,\alpha)$ for fixed value of the temperature $t_0=0.05$ and
    asymmetry parameter $\varepsilon_0=1/3$.}
    \label{Rppp}
    \end{center}
\end{figure}
\begin{figure}[t]
    \begin{center}
    \epsfig{file=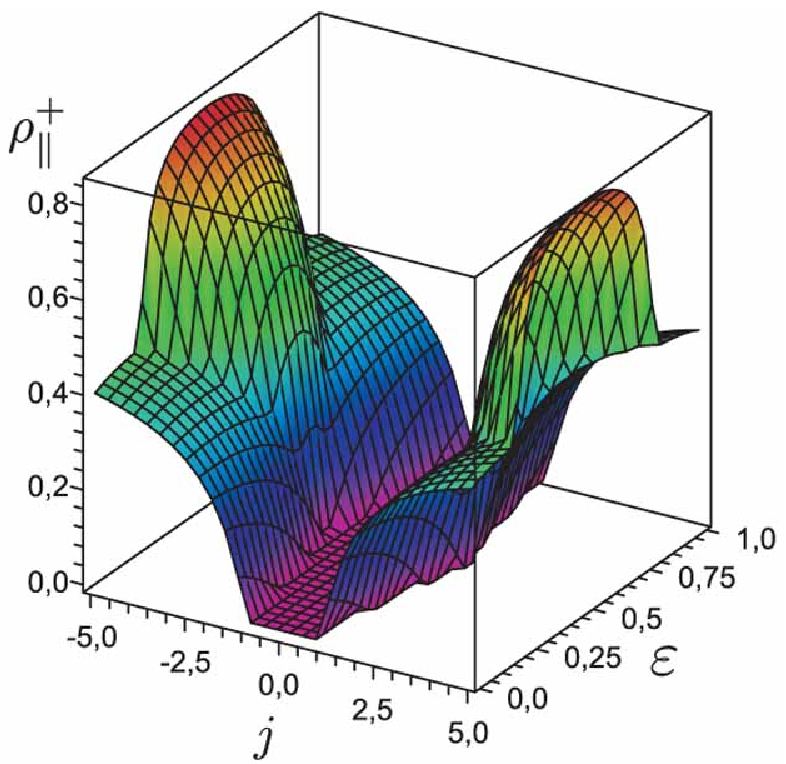,width=0.4 \textwidth}
    \caption{The dependence $\rho^+_\parallel(j,\varepsilon)$ for fixed value of the temperature $t_0=0.05$
    and angle $\alpha_0=0$.}
    \label{Rpppt}
    \end{center}
\end{figure}
\begin{figure}[t]
    \begin{center}
    \epsfig{file=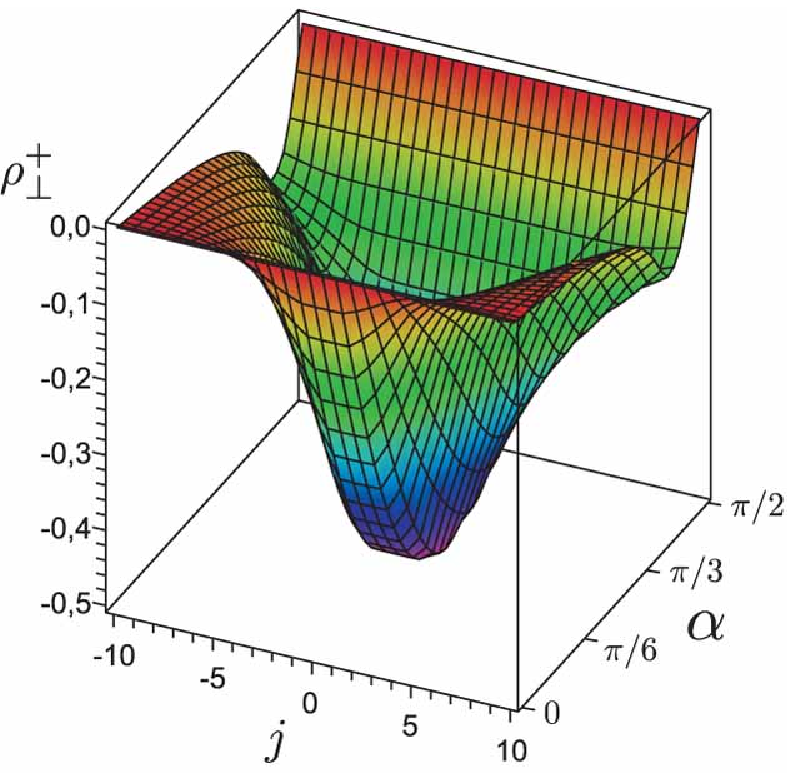,width=0.4 \textwidth}
    \caption{The dependence $\rho^+_\perp(j,\alpha)$ for fixed value of the temperature $t_0=0.05$ and
    asymmetry parameter $\varepsilon_0=1/3$.}
    \label{Rppr}
    \end{center}
\end{figure}
\begin{figure}[t]
    \begin{center}
    \epsfig{file=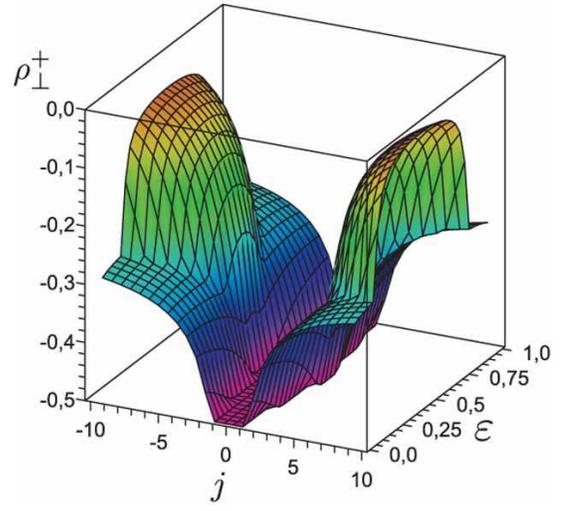,width=0.4 \textwidth}
    \caption{The dependence $\rho^+_\perp(j,\varepsilon)$ for fixed value of the temperature $t_0=0.05$
    and angle $\alpha_0=\pi/4$.}
    \label{Rpprt}
    \end{center}
\end{figure}

The limiting values of the qualitatively similar dependences
$\rho^+_\parallel(j_y)$ and $\rho^+_\parallel(t)$ corresponding to
the TAFF regime of vortex motion transverse to the pinning
channels are determined by guided vortex motion along the pinning
channels and grow with increasing magnitude of the angle $\alpha$
since in this case the component of the Lorentz force along the
pinning channels increases. In the FF regime, as the pinning
viscosity becomes isotropic the contribution to the dependences
$\rho^+_\parallel(j)$ and $\rho^+_\parallel(t)$ due to vortex
motion transverse to the PPP channels becomes substantial, and the
limiting values of these dependences are equal to unity (see Figs.
$\ref{Rppp}$ - \ref{Rpprt}).

The main contribution to the even transverse resistivity
$\rho^+_\perp$ is proportional to the factor $\sin(2\alpha)/2$;
therefore, the angle most favorable for its observation is
$\alpha=\pi/4$. The current dependence $\rho^+_\perp(j)$ and the
temperature dependence $\rho^+_\perp(t)$ have their maximum
absolute values in the TAFF regime of vortex motion transverse to
the PPP channels (the same value is approached if the angle is
replaced by its complement in the limit $j \to 0$ and $t \to 0$)
and go to zero with the onset of the FF regime as a consequence of
isotropization of the pinning viscosity (see Fig. \ref{Rppr}). The
resistivity $\rho^+_\perp$ can serve as a measure of the
anisotropy of the pinning viscosity since it is determined by the
difference of the pinning viscosities transverse to and along the
pinning channels (see also Eqs. \ref{f28}, \ref{f29}).

As can be seen from Figs. \ref{Vp3d}, \ref{Rpppt} and \ref{Rpprt},
that behavior of the $\rho^+_{\parallel,\perp}(j,\varepsilon)$
resistivities is closely equal to the behavior of the
$\nu^+(f,\varepsilon)$. It is also follows from Eqs. (\ref{f28}),
(\ref{f29}), that all that had told about behavior of the
$\nu^+(f,\varepsilon)$ function can be repeated by analogy here.
Hence, the step-like appearance of the
$\rho^+_{\parallel,\perp}(j)$ is direct consequence the asymmetry
of the pinning potential. A unique distinction is the influence of
the angle $\alpha$ on the
$\rho^+_{\parallel,\perp}(j,\varepsilon)$ resistivities by means
of internal angle dependence. The internal angle dependence  is
reduces current's influence to $\nu^+(j)$ function and cause to
expansion of the even resistivities along $j$ axis when $\alpha$
increases.

\begin{figure}[th]
    \begin{center}
    \epsfig{file=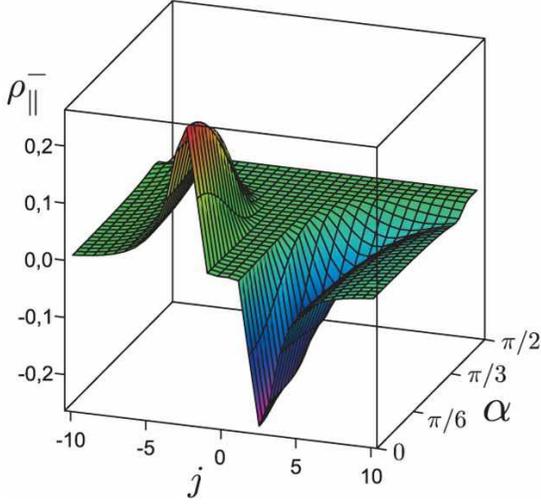,width=0.4 \textwidth}
    \caption{The dependence $\rho^-_\parallel(j,\alpha)$ for fixed value of the temperature $t_0=0.05$ and
    asymmetry parameter $\varepsilon_0=1/3$.}
    \label{Rmpp}
    \end{center}
\end{figure}

\begin{figure}[th]
    \begin{center}
    \epsfig{file=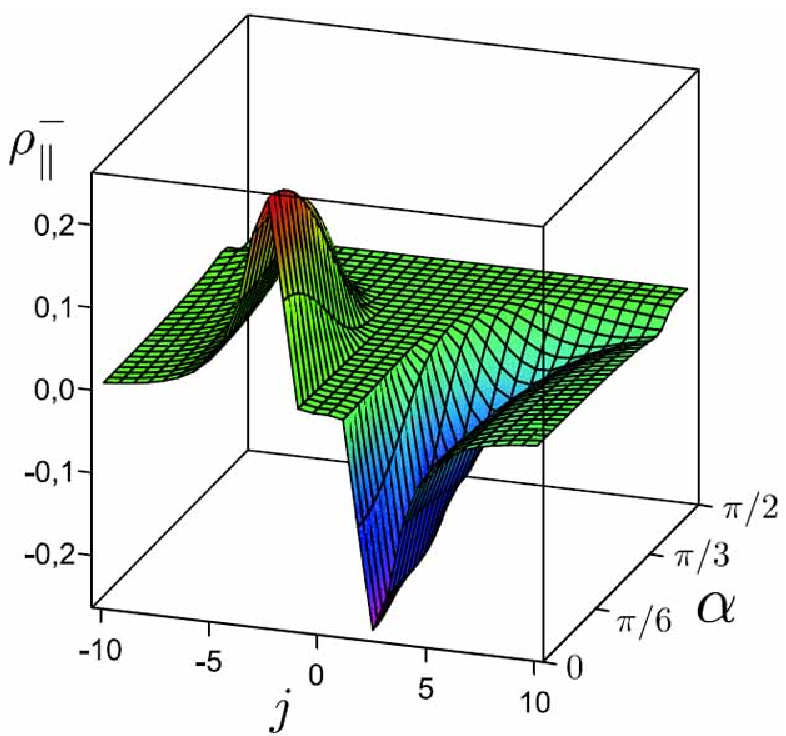,width=0.4 \textwidth}
    \caption{The dependence $\rho^-_\parallel(j,\varepsilon)$ for fixed value of the temperature $t_0=0.05$ and
    angle $\alpha_0=0$.}
    \label{Rmppt}
    \end{center}
\end{figure}

\begin{figure}[th]
    \begin{center}
    \epsfig{file=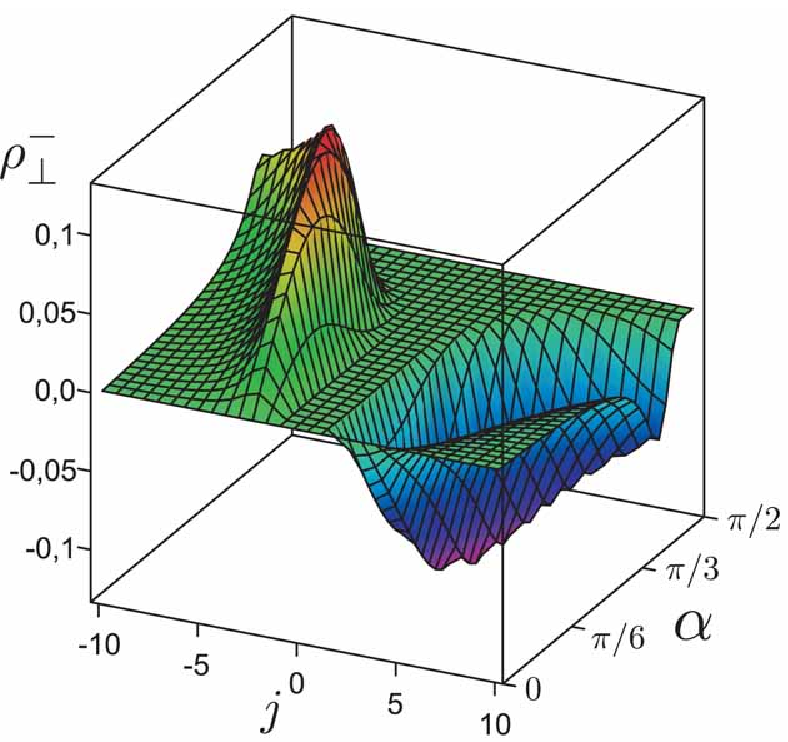,width=0.4 \textwidth}
    \caption{The dependence $\rho^-_\perp(j,\alpha)$ for fixed value of the temperature $t_0=0.05$
    and asymmetry parameter $\varepsilon_0=1/3$.}
    \label{Rmpr}
    \end{center}
\end{figure}

\begin{figure}[th]
    \begin{center}
    \epsfig{file=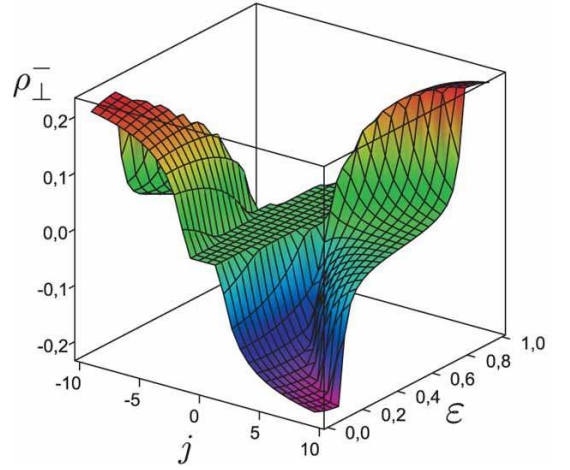,width=0.4 \textwidth}
    \caption{The dependence $\rho^-_\perp(j,\varepsilon)$ for fixed value of the temperature $t_0=0.05$
    and angle $\alpha_0=\pi/4$.}
    \label{Rmprt}
    \end{center}
\end{figure}

As was noted above, the odd longitudinal $\rho^-_\parallel$ and
transverse $\rho^-_\perp$ magnetoresistivities arise thanks to the
asymmetry of the pinning potential, and therefore their
characteristic scale is proportional to $\nu^-$ (see Eqs.
\ref{f30}, \ref{f31}). Therefore, their qualitative form is
inherited completely by the behavior of $\nu^-$ as a function of
the current, asymmetry parameter and temperature (Figs. \ref{Rmpp}
- \ref{Rmprt}).


A characteristic peak appears in the dependencies
$\rho^-_\parallel(j)$ in the region of nonlinearity of $\nu^-$ as
a function of the current and parameter of asymmetry while in the
TAFF and FF regimes of vortex motion transverse to the pinning
channels they vanish (Figs. \ref{Rmpp}, \ref{Rmppt}). The
temperature behavior of the resistivities $\rho^-_\perp$ and
$\rho^-_\parallel$ is similar to $\nu^-(t)$ (see Figs. \ref{Vm3d},
\ref{Vmet3d}). As the main contribution to the odd transverse
resistivity $\rho^-_{\perp}$ is proportional to the factor
$\sin(2\alpha)/2$, then the angle most favorable for its
observation is $\alpha=\pi/4$. It can be important for experiment,
that the maximal value of the resistivity $\rho^\pm_\perp$ does
not exceed $1/2$, as it follows from Eqs. (\ref{f29}),
(\ref{f30}).

As was noted above, the resistivity internally depends from the
angle $\alpha$ such as $f=j_y=j\cos\alpha$, and it follows from
this, that value of the transport current density, when
resistivity $\rho^-_\perp$ is maximal, will be expressed as:
\begin{equation}
\label{Jmax}
    j_{max}=\min(j_{cr1},j_{cr2})/\cos\alpha.
\end{equation}
If $\alpha$ tends to $\pi/2$, when $j_{max}$ tend to infinity. It
physically means, that the Lorentz force, which is acting to the
vortices, there is parallel to the pinning channels and can't
drags the vortices across the pinning channels.

It is worth noticing, that the $j_{cr1}$ and $j_{cr2}$ are
functions of the asymmetry parameter $\varepsilon$ as it was
proved in Section \ref{discussion}. This explains the fact, that
if $\varepsilon \rightarrow 0,1$ the $\rho^-_{\parallel,\perp}$
tends to step-like or to bump-like form. It happens because one of
the pinning forces is becoming infinite.

\subsection{The angular stability of the resistivities in LT - geometries.}\label{Mrstabil}

Let us consider the observed resistivities in the T and L
geometries, where the current is directed exactly parallel
($\alpha=0$) or perpendicular ($\alpha=\pi/2$) to the PPP
channels. It follows from formulas (\ref{f28}) - (\ref{f31}) that
in these limiting cases $\rho^\pm_\perp=0$, and we obtain for
$\rho^+_\parallel$ and $\rho^-_\parallel$
\begin{equation}
\label{f39}
    \rho^+_{\parallel,\textnormal{T}}=\nu^+_\textnormal{T}, \quad
    \rho^-_{\parallel,\textnormal{T}}=\nu^-_\textnormal{T}
    \quad (\alpha=0, \textnormal{T geometry}),
\end{equation}
\begin{equation}
\label{f39b}
    \rho^+_{\parallel,\textnormal{L}}=1, \quad \rho^-_{\parallel,\textnormal{L}}=0
    \quad (\alpha=\pi/2, \textnormal{L geometry}),
\end{equation}
where longitudinal even $\rho^+_{\parallel,\textnormal T}$ and odd
$\rho^-_{\parallel, \textnormal T}$ resistivities are due to
vortex motion transverse to the PPP channels, and described by the
functions $\nu^+_\textnormal{T}=\nu^+(j,t,\varepsilon)$ and
$\nu^-_\textnormal{T}=\nu^-(j,t,\varepsilon)$ respectively. In the
limit $j,t \to \infty$ we have
$\rho^+_{\parallel,\textnormal{T}}=1$,
$\rho^-_{\parallel,\textnormal{T}}=0$. The resistivity
$\rho^+_{\parallel, \textnormal L}$ in the L geometry is equal to
unity due to guided vortex motion along the PPP channels, for
which pinning is absent.

Formula (\ref{f39}) expresses simple relations between the
observable resistivities $\rho^+_{\parallel, \textnormal T}$ and
$\rho^-_{\parallel, \textnormal T}$ in the T geometry. The form of
the functions $\nu^{\pm}_\textnormal{T}$ can be reconstructed, as
can be seen from formulae (\ref{f39}), from the measurements of
$\rho^+_{\parallel,\textnormal T}$ and
$\rho^-_{\parallel,\textnormal T}$.

Therefore, it makes sense to consider the question of the
stability of the measurements in these geometries since the
preparation of the samples can lead to small deviations
$\delta\alpha$ from the values $\alpha=0,\pi/2$. Here it should
also be borne in mind that besides the resistivities
$\rho^+_\parallel$, and $\rho^-_\parallel$ assigned by formulas
(\ref{f28}) and (\ref{f31}), in the presence of an angle
deviation, $\alpha$, the resistivities $\rho^+_\perp$ and
$\rho^-_\perp$, not present in the L and T geometries, also
appear. The expansions of $\rho^\pm_{\parallel,\perp}$ in $\alpha$
about $\alpha=0$ (in the T geometry) and in
$\Delta\alpha=\pi/2-\alpha$ about $\alpha=\pi/2$ (in the L
geometry) out to the first nonvanishing terms have the form:
\begin{equation}
\label{f40}
    \rho^-_{\perp,\textnormal{T}}=\nu^-_\textnormal{T}(j)\alpha + o(\alpha^3),
\end{equation}
\begin{equation}
\label{f41}
    \rho^-_{\parallel,\textnormal{T}}=\nu^-_\textnormal{T}(j)-\left(\frac12\frac{\partial\nu^-_\textnormal{T}(j)}{\partial j}j+
    \nu^-_\textnormal{T}(j)\right)\alpha^2 + o(\alpha^3),
\end{equation}
\begin{equation}
\label{f42}
    \rho^+_{\perp,\textnormal{T}}=(\nu^+_\textnormal{T}(j)-1)\alpha + o(\alpha^3),
\end{equation}
\begin{equation}
\label{f43}
    \rho^+_{\parallel,\textnormal{T}}=\nu^+_\textnormal{T}(j)+
    \left(1-\frac12\frac{\partial\nu^+_\textnormal{T}(j)}{\partial j}j-\nu^+_\textnormal{T}(j)\right)\alpha^2 + o(\alpha^3),
\end{equation}
\begin{equation}
\label{f44}
    \rho^-_{\perp,\textnormal{L}}=\left.-\frac{\partial\nu^-(j)}{\partial j}\right|_{j=0}j(\Delta\alpha)^3+o((\Delta\alpha)^4),
\end{equation}
\begin{equation}
\label{f45}
    \rho^-_{\parallel,\textnormal{L}}=\left.\frac{\partial\nu^-(j)}{\partial j}\right|_{j=0}j(\Delta\alpha)^3 + o((\Delta\alpha)^4),
\end{equation}
\begin{equation}
\label{f46}
    \rho^+_{\perp,\textnormal{L}}=(\nu^+(0)-1)(\Delta\alpha) + o((\Delta\alpha)^2),
\end{equation}
\begin{equation}
\label{f47}
    \rho^+_{\parallel,\textnormal{L}}=1+(\nu^+(0)-1)(\Delta\alpha)^2+o((\Delta\alpha)^3).
\end{equation}

Below we will use simple physical arguments in order to estimate a
value and to explain all main features of the resistivities
(\ref{f40})-(\ref{f47}). The main cause of the presented behavior
of the resistivities in the L geometry is extremely small inner
dependence ($f \approx j\alpha$) of the $\nu^\pm$ functions from
the transport current density. Besides, it is easy to see, that
resistivities $\rho^-_{\perp,\textnormal{L}}$ and
$\rho^-_{\parallel,\textnormal{L}}$ are close to zero for $t\ll1$
and $j\ll1$. It happens because the derivative of the $\nu^-$ is
nonzero only in the vicinity of transition from the full guiding
regime to the TAFF regime and from the TAFF to the FF regime. If
temperature and current will be rise, it causes the derivate and
appropriate resistivities grows until the FF regime does not
happen. On the other hand, the resistivity
$\rho^+_{\perp,\textnormal{L}}$ in the L geometry varies linearly
for small deviations of the $\alpha$ and does not depend on the
current density. In the same way the
$\rho^+_{\parallel,\textnormal{L}}$ does not depend on the small
deviation of $\alpha$ and on the current density.

In the T geometry the inner dependence of the $\nu^\pm$ from the
current density is strong ($f \approx j$). The resistivities
$\rho^+_{\parallel,\textnormal T}$ and
$\rho^-_{\parallel,\textnormal T}$ depend only on $\nu^+$ and
$\nu^-$ functions and have a weak angle dependence accordingly.
The resistivities $\rho^-_{\perp,\textnormal T}$ and
$\rho^+_{\perp,\textnormal T}$ are proportional to $\alpha$
deviation. Similarly to foregoing we can conclude, that the
resistivity $\rho^\pm_{\perp,\textnormal{T}}$ will be more
unstable in comparison with $\rho^\pm_{\parallel,\textnormal{L}}$
for a small deviation of the angle $\alpha$ from the T geometry.

The relative deviation of the resistivity for a small deviation
from the T and L geometries for $\rho^+_\parallel$ it is of the
order $\Delta\rho^+_{\parallel,\textnormal
T}/\rho^+_{\parallel,\textnormal T} \sim \alpha^2/\nu(j,t)$ in the
T geometry and $\Delta\rho^+_{\parallel,\textnormal
L}/\rho^+_{\parallel,\textnormal L} \sim \Delta\alpha$ in the L
geometry. Thus, $\rho^+_{\parallel,\textnormal T}$ is the most
unstable in the TAFF regime of vortex motion transverse to the
pinning channels, where $\nu(j,t)\ll 1$. The physical reason for
this behavior is the rapid variation of the angle $\beta$ from
$\alpha=0$ in the T geometry, where $v_y=0$, to the angle
corresponding to the guiding regime with $v_y \gg v_x$.

The behavior of the resistivities in the L geometry is physically
clear from the fact that for $\alpha \simeq \pi/2$ the angle
$\beta$ varies hardly at all, i.e., the direction of the velocity
vector $\mathbf v$ varies only slightly (in contrast to the case
of the T geometry) and thermally activated transitions of the
vortices through pinning potential barriers play main role here.

As was stated above, in an actual experiment small deviations of
the angle $\alpha$ from the values $\alpha=0,\pi/2$ corresponding
to the L and T geometries are always present. Utilizing
experimental measurements of $\rho^\pm_{\parallel,\perp}$, these
deviations can be found using the following scheme. First,
neglecting small quadratic contributions in $\alpha$ to the
resistivities $\rho^-_{\parallel,\textnormal{T}}$ and
$\rho^+_{\parallel,\textnormal{T}}$ (in the region where they are
stable), it is possible to solve the inverse problem using
formulas (\ref{f41}) and (\ref{f43}), i.e., to reconstruct the
function $\nu$. Knowing this, from the formulas for the
resistivity $\rho^+_{\perp,\textnormal{L}}$, which vanish in the L
geometry and are linear for small deviations $\alpha$, it is
possible to find the corresponding value of $\alpha$ deviations.
The self-consistency of this scheme is checked by calculating the
quadratic corrections in $\alpha$ and $\Delta\alpha$, which should
be small relative to the main contribution in the T and L
geometries.

\subsection{Weak and strong asymmetry.}\label{weakassym}

Let us discuss firstly the case when asymmetry of the pinning
potential is very small, i.e.
\begin{equation}
\label{f48}
    \varepsilon=1/2+z,
\end{equation}
where $z \to 0$ is the small deviation of the asymmetry parameter
from the symmetric case. Substituting Eq.(\ref{f48}) into Eq.
(\ref{f20}) we can expand the $\nu(f,t,\varepsilon)$ in a Taylor
series about small deviation $z$ up to the second-order terms. A
convenient result can be presented in the following form:
\begin{equation}
\label{f49}
    \nu \approx \tilde{\nu} = \tilde{\nu}^+ + \tilde{\nu}^-,
\end{equation}
where
\begin{equation}
\label{f50}
    \tilde{\nu}^+=(f^2-4)^2/(f^2(f^2-4)-G),
\end{equation}
corresponds to the even component of the $\tilde{\nu}$ function
expansion into a Taylor series, whereas
\begin{equation}
\label{f51}
   \tilde{\nu}^- = Wz,
\end{equation}
corresponds to the odd component of the $\tilde{\nu}$, \\
$G=16ft(\cosh(f/(2t))-\cosh(1/t))/\sinh(f/(2t))$, and \\
$W=16(4-f^2)(G+(4-f^2)(f\sinh(1/t)/\sinh(f/(2t))+2)/(f(4-f^2)+G)^2)$.

Notice now, that $\tilde{\nu}^+$ function in Eq. (\ref{f50}) is
the even function of the external motive force $f$ and coincides
with the similar expression given by Eq. (\ref{f27}) which earlier
was pointed out in \cite{UsSh}. It is easy also to prove, that
$\tilde{\nu}^- $ in Eq. (\ref{f51}) is odd with respect to the $f$
and $z$ respectively.

From Eqs. (\ref{f51}), (\ref{f30}) and (\ref{f31}) we can
calculate an expression for the asymmetry parameter $\varepsilon$:
\begin{equation}
\label{f52}
    \varepsilon=1/2+\rho^-_{\parallel,\textnormal T}/W.
\end{equation}
Note that Eq. (\ref{f52}) can be used for calculating the value of
the $\varepsilon$ from the experimental data in the limit
$\alpha\to0$ and $\varepsilon\approx 1/2$.

In the opposite case, when a pinning force in one direction is
considerably larger than a pinning force in another direction, the
strong asymmetry arises. Let us consider the case where
$\varepsilon$ is the small deviation of the asymmetry parameter in
the strong asymmetry case (when $f_{cr1}=-1$ and
$f_{cr2}=+\infty$). Similarly to the weak asymmetry case, we
expand the $\nu(f,t,\varepsilon)$ in Taylor series about small
deviation $\varepsilon$ up to second-order terms:
\begin{equation}
\label{f54}
    \nu(f,t,\varepsilon)=R_0(f,t)+R_1(f,t)\varepsilon,
\end{equation}
where
\begin{equation}
\label{f55}
    R_0=(f+1)^2/(fX),
\end{equation}
\begin{equation}
\label{f56}
    R_1=C/X^2,
\end{equation}
$X=1+f+2t\sinh(1/(2t))\sinh((1+f)/(2t))/\sinh(f/(2t))$, \\
$C=(4f^2+5f+2)/f+(
(2f^2(t\cosh(f/(2t))-t\cosh((f+2)/2t)+\cosh(1/(2t))\cosh((f+1)/(2t)))+
\sinh((2+f)/(2t))(1+2f)) - 4t\sinh(1/(2t))\sinh((f+1)/(2t))(3f+2)
 )/\sinh(f/(2t))$.

 The $R_0$ and $R_1$ dependences neither even or odd functions.
 It follows from this, that $\nu^+$ and $\nu^-$ functions depend
 on the small deviation $\varepsilon$ of the asymmetry parameter and will be
 able to present as $\nu^+=G_1+G_2\varepsilon$, $\nu^-=H_1+H_2\varepsilon$, where $G_{1,2}$
 is the even part of the $R_{1,2}$ functions and $H_{1,2}$ is the odd part
  of the $R_{1,2}$ functions respectively. This fact can help to
  extract the $\varepsilon$ parameter from experiential data as:
\begin{equation}
\label{f57}
    \varepsilon=(\rho^+_\parallel-G_1)/G_2.
\end{equation}

Note that Eq. (\ref{f57}) can be used for calculating the value of
the $\varepsilon$ from the experimental data in the limit
$\alpha\to0$ and $\varepsilon\approx 0$ (the case
$\varepsilon\approx 1$ it is possible to consider by substitution
$\varepsilon$ to $1$ and $f$ to $-f$ in formulas
(\ref{f54})-(\ref{f57})).

\section{Conclusion.}\label{end}
In this work we proposed exactly solvable \cite{ShSo} two-
dimensional model structure for study of the ratchet effect in
superconducting film in presence of the asymmetric planar pinning
potential as was studied by experiment firstly in \cite{Morrison}.

We have theoretically examined the strongly nonlinear resistive
behavior of the two-dimensional vortex system of a superconductor
as a function of the transport current density $\mathbf j$, the
temperature $t$, and the angle $\alpha$ between the directions of
the current and the PPP channels. The nonlinear (in $\mathbf j$)
resistive behavior of the anisotropic vortex ensemble can be
caused by factor of  a "pinning" origin which takes into account
the presence of  anisotropic pinning with asymmetry of the PPP. It
is physically obvious that such pinning at low enough temperatures
leads to anisotropy of the vortex dynamics since it is much easier
for vortices to move along the pinning channels (the guiding
effect in the flux-flow regime, which is linear in the current)
than in the perpendicular direction, where it is necessary for
them to overcome the pinning potential barriers from the pinning
channels, which also is a source of resistive nonlinearity. If
under variation of one of the "external" parameters $\mathbf j$,
$t$, $\alpha$ the intensity of manifestation of the indicated
nonlinearity is weakened, then this weakening will lead to an
"effective isotropization" of the vortex dynamics, i.e., to a
convergence (and in the limit of the absence of nonlinearity , to
coincidence) of the directions of the mean velocity vector of the
vortices and the Lorentz force.

It is physically clear that the current, temperature, and angle
$\alpha$ have a qualitatively different effect on the weakening of
pinning  and the corresponding transition from anisotropic vortex
dynamics to isotropic. With growth of $\mathbf j$ the Lorentz
force $\mathbf F_{\textnormal L}$ grows and the height of the
potential barrier decreases, so that for $j \geqslant
j_{cr1},j_{cr2}$ (where $j_{cr1},j_{cr2}$ are the crossover
currents of the indicated transition, whose width grows with
growth of $t$) this barrier essentially disappears. The quantities
$j_{cr1},j_{cr2}$ depend on $\alpha$ by virtue of the fact that
the probability of overcoming the barrier is governed not by the
magnitude of the force $F_\textnormal L$, but only by its
transverse component $F_\textnormal L\cos\alpha$, so that
$j_{cr1,2}(\alpha) =j_{cr1,2}(0)/\cos\alpha$ grows with growth of
$\alpha$. Since an increase in the temperature $t$ always
increases the probability of overcoming the pinning barrier, the
transition to isotropization of the vortex dynamics is that much
steeper in $t$, the smaller is the pinning barrier.

In order to theoretically analyze the above-described physical
picture of a nonlinear anisotropic resistive response, Sections
\ref{FPmetod} and \ref{APPP} employed a comparatively simple, but
at the same time quite realistic, planar model of stochastic
pinning. It allows one to reduce the calculations to the
evaluation of analytical formulas (\ref{f28})-(\ref{f31}), which
have a simple physical interpretation. A distinguishing feature of
this model is the possibility, within the framework of a unified
approach, to describe consistently the nonlinear transition from
the anisotropic dynamics of a vortex system (for currents $j \ll
j_{cr1,2}(\alpha)$ at relatively low temperatures) to isotropic
behavior (for currents $j > j_{cr1,2}(\alpha)$ at relatively high
temperatures). In the model under consideration this approach
corresponds (for $t>0$) to a substantially nonlinear crossover
from the linear low-temperature thermally activated flux-flow
regime to the ohmic flux-flow regime of vortex motion.

Proceeding now to a brief description of the main theoretical
results, we note here that an analytical representation of the
nonlinear resistive response of the investigated system in terms
only of elementary functions was possible thanks to the use of a
simple but physically realistic model of anisotropic pinning with
asymmetric sawtooth PPP (see Sec. \ref{APPP} and Fig.
\ref{Potential}). The exact solution obtained made it possible for
the first time to consistently analyze not only the qualitatively
clear dynamics of the nonlinear guiding effect, but also the
nontrivial question of the interaction of guided vortex motion
along PPP channels and the ratchet effect. The most important
result in our opinion is the conclusion that the appearance of
novel $\rho^-_{\parallel,\perp}$ magnetoresistivities does not
require (as it was in (\cite{ShSo})) the Hall effect (see Sec.
\ref{APPP}). The nonlinear formulas (\ref{f30}) and (\ref{f31}) in
agreement with physical intuition (now already nonlinear) clearly
demonstrate that the most natural and "sufficient" reason for the
relatively large novel $\rho^-_{\parallel,\perp}$ - effects is the
asymmetry of the pinning wells. At comparatively low temperatures
and weak currents it leads to the realization of a quite intense
(over a wide interval of angles around $\alpha=\pi/4$) guided
vortex motion along the pinning channels in the thermally
activated flux-flow regime, i.e., to the appearance of
$\rho^+_\perp$-effects, and at currents $j \approx
j_{cr1,2}(\alpha)$, to the appearance of characteristic maxima in
the curves of the odd components of the resistivities
$\rho^-_{\parallel,\perp}$ (see Subsection \ref{Rcharact} and
Figs. \ref{Rmpp}-\ref{Rmprt} ).

A completely novel result of the present work is also contained in
formulas (\ref{f30}) and (\ref{f31}). It is a quantitative
description of the interaction of the guiding effect and the
ratchet effect, which is valid for all possible values of the
asymmetry parameter $0<\varepsilon<1$ . Formally, this interaction
arises as a result of the fact that in the case of anisotropic
pinning on asymmetric PPP the force of the overcoming the pinning
well (see Eq. (\ref{f19})), which determines the probability of
overcoming the potential barrier (and therewith also determines
the magnitude of the component of the vortex velocity
perpendicular to the pinning channels), is different in the
opposite directions of the $x$-axis. Then arising of the odd
resistivities defined by Eqs. (\ref{f30}), (\ref{f31}), appears
only due to the ratchet form of the PPP and to the change of their
sign with the current or magnetic field reversals (see Eq.
\ref{f21}). Their origin follows from the emergence of a certain
equivalence of the $xy$-direction for the case, that a guiding of
vortices along the channels of the washboard PPP is realized at
$\alpha \neq 0,\pi/2$. Note also that for $\alpha=0$ Eq.
(\ref{f31}) gives in fact the ratchet signal measured in
\cite{Villegas}. The key point in the physical interpretation of
these formulas is our treatment of the function
$\nu(f,t,\varepsilon)$ as the probability of overcoming the
potential barrier of the pinning channel, from which follows an
understanding of the evolution of the functions associated with
it, $\nu^\pm$ (see Subsection \ref{APPP}), as functions of the
magnitude of the current density $j$, temperature $t$, and angle
$\alpha$. Note that this treatment is not a unique property of the
stochastic model of anisotropic pinning considered in this work,
but can also be consistently realized within the framework of the
nonlinear phenomenological approach under much broader
assumptions, including, in particular, an account of the
inter-vortex interaction.

If, as is usually the case in experiment \cite{Morrison}, that the
asymmetry of the pinning potential are sufficiently small
($\varepsilon \approx 1/2$), then formulas (\ref{f28})-(\ref{f31})
simplify substantially since under these conditions $\nu^- \sim
(1/2+z), \quad z \to 0$ (see Subsection \ref{weakassym}).

In conclusion, it should also be noted that ratchet effect opens
up the possibility for a variety of experimental studies of
directed motion of vortices simply by measuring longitudinal and
transverse voltages. Experimental control of amplitude and
frequency of the external force, damping, anisotropy parameters,
and temperature can be easily provided. In contradiction with
other vortex-based ratchet models, the one presented here allows
to separate the Hall and ratchet voltages which are similar in
their $(j,t)$ behavior, but have different origin and magnitude.
Note also that the new ratchet voltages disappear during the
procedure of the "current averaging" frequently used in
experiments \cite{Soroka} for the cancelation of parasitic
thermoelectric voltages.

\end{document}